\documentclass[twocolumn,showpacs,amsmath,amssymb,prl,longbibliography]{revtex4}

\usepackage{natbib}
\usepackage{amsmath} 
\usepackage{amsfonts} 
\usepackage{color}
\usepackage{nicefrac}
\usepackage{dsfont}
\usepackage{graphicx}
\usepackage[all,2cell]{xy}
\usepackage{mathtools}

\DeclarePairedDelimiter\floor{\lfloor}{\rfloor}

\newcommand*{\Scale}[2][4]{\scalebox{#1}{$#2$}}

\begin{document}

\title{Tight uncertainty relations for cycle currents}

\author{Matteo Polettini}

\email{matteo.polettini@uni.lu}
\affiliation{Department of Physics and Materials Science, University of Luxembourg, Campus
Limpertsberg, 162a avenue de la Fa\"iencerie, L-1511 Luxembourg (G. D. Luxembourg)} 

\author{Gianmaria Falasco}
\affiliation{Department of Physics and Materials Science, University of Luxembourg, Campus
Limpertsberg, 162a avenue de la Fa\"iencerie, L-1511 Luxembourg (G. D. Luxembourg)} 
 
\author{Massimiliano Esposito}
\affiliation{Department of Physics and Materials Science, University of Luxembourg, Campus
Limpertsberg, 162a avenue de la Fa\"iencerie, L-1511 Luxembourg (G. D. Luxembourg)} 

\date{\today}

\begin{abstract}
Several recent inequalities bound the precision of a current -- counting net number of transitions in a system -- by a thermodynamic measure of dissipation. However, while currents may be defined locally, dissipation is a global property. Inspired by the fact that ever since Carnot cycles are the unit elements of thermodynamic processes, we prove similar bounds tailored to cycle currents -- counting net cycle completions -- in terms of their conjugate affinities. We show that these inequalities are stricter than previous ones, even far from equilibrium, and that they allow to tighten those on transition currents. We illustrate our results with a simple model and discuss some technical and conceptual issues related to shifting attention from transition to cycle observables.
\end{abstract} 



\maketitle

In recent years several variants of a thermodynamic uncertainty relation (TUR) have been derived, bounding the precision of an observable by a quantity of clear physical interpretation. In particular, one-half the mean entropy flow rate $\sigma$ is an upper bound to the squared-signal-to-noise ratio of a stationary thermodynamic current $\varphi_a$.  In other words, precision costs: the more precise the current, the more the dissipation. In formula we can cast this as a bound on a current's dispersion:
\begin{align}
\overline{\mathfrak{d}}_a : =  \frac{\overline{\kappa}^{(2)}_a}{ | \overline{\kappa}^{(1)}_a |}  \geq   \frac{2  }{ \sigma /  | \overline{\kappa}^{(1)}_a|},
 \label{eq:TUR}
\end{align}
where $\overline{\kappa}^{(1)}_a$ and $\overline{\kappa}^{(2)}_a$  are the current's mean and variance, and the overline signals that cumulants are estimated and scaled over long times.

A common framework to prove these results is that of discrete-state space, continuous-time stationary Markov walks (CTSMW) \cite{bar15, pie16, gin16, pie17, gin17, hor17}. Other derivations encompass periodic states \cite{pro17} and relaxation \cite{pie17, dec18, has19, dec20}, possibly non-Markovian and subject to feedback, as well as time-symmetric observables and first-passage times \cite{gar17, gin17b, dit18}. Large deviation and information theory allow unified formulations: in particular TURs for observables that are odd under an involution (e.g. time-reversal) follow from the Hilbert structure of the space of observables \cite{fal20}. TURs are the more meaningful the tighter: the bound Eq.\,(\ref{eq:TUR}) saturates close to equilibrium only if the current is the entropy flow itself, which is a global observable defined over the entire state space \cite{pol16}. 
\begin{figure}[t!]
\begin{center}
\includegraphics[width=\columnwidth]{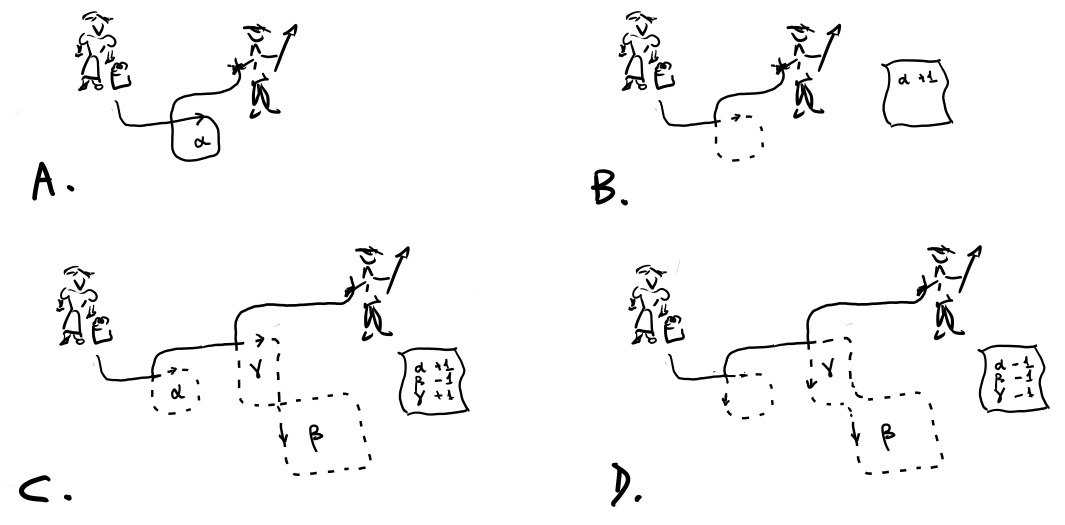}
\caption{{ A) The first cycle $\alpha$ walked by Theseus in clockwise direction; B) the strand removed and the cycle recorded; C) more cycles recorded: notice that $\beta$ is in anti-clockwise direction and that $\gamma$ is started before $\beta$ but completed after; D) a partial inversion of cycles $\alpha$ and $\beta$ on Theseus's way back.}}
\label{fig:dedalus}
\end{center}
\end{figure}

Pursuing a line of research that aims at casting global results local \cite{pol19,bis17,mar19}, in this manuscript we show how to produce tighter bounds on the currents. The key insight is to shift attention from transition currents $a = x'x$ (counting net transitions from a state  $x$ to another $x'$) to cycle currents $a = c$ (from a state back to itself via cycle $c$). One of several possible procedures to define a set of cycle currents along a realization of a CTSMW is illustrated in Fig.\,\ref{fig:dedalus}, and can be told in terms of an ancient Greek myth. Suppose the Markov walker is Theseus, wandering around the Knossos labyrinth. As Theseus proceeds he lays Ariadne's thread so that, once the Minotaur is found and killed, he will be able to trace his steps back to the entrance. However, in our thermodynamic twist of the story, whenever Theseus accidentally encounters the filament he laid, 
 he cuts it, wraps up the thread behind him and sews the strands' ends together, annotating the cycle he performed (but, because he is Markovian, later on Theseus may traverse the same cycle again). Cycle currents are the net number of times a cycle without crossings is performed as listed in Theseus's parchment with respect to some orientation (e.g. clockwise/anticlockwise). Our main result is then
\begin{align}
\mathfrak{d}_c(t) \geq \frac{2}{F_c}, \label{eq:cyclebound}
\end{align}
where $F_c$ is the so-called cycle affinity and here estimates entering in the dispersion, without overline, are  istantaneous, at any given time, and with respect to any distribution of the starting state.

As a second result, we tighten the stationary bound Eq.\,(\ref{eq:TUR}) for transition currents by replacing $\sigma$ with $\sigma_{x'x} \leq \sigma$, a reduced measure of the entropy flow rate along all cycles that contain transition $x'x$. Finally, we provide some computational evidence for { the long-time analog of Eq.\,(\ref{eq:cyclebound}),  $\overline{\mathfrak{d}}_c \geq 2/F_c $}. However, a proof of this latter relation remains elusive because of the non-additive nature of cycle currents, which are a different kind of observable with respect to transition currents and pose interesting and specific challenges.

\paragraph{Setup.} Thermodynamics deals with time-integrated currents $\phi_a(t_0,t)$ measured in an interval $[t_0,t]$. Currents are powered by conjugate forces $F_a$; without loss of generality we let all $F_a >0$. The entropy flow $\Sigma := \sum_a F_a \, \phi_a$ quantifies dissipation, and by ``thermodynamic consistency'' we mean that all representations in terms of different notions of current lead to the same entropy flow function, up to boundary terms.

In the stochastic framework currents are random variables, functionals $\phi_a(t_0,t) = \phi_a[\omega]$ of stochastic trajectories $\omega$ which we assume to be a CTSMW on state space $X \owns x$, with time-independent rates $r_{x'x} > 0$ of jumping from $x$ to $x'$. A trajectory is a succession of visited states $x_i$ and soujourn times $\tau_i$ up to total time $\sum_{i=0}^{n} \tau_i = t-t_0$,
\begin{align}
\omega = (x_0,\tau_0) \to (x_1,\tau_1) \to \ldots  \to (x_{n},\tau_{n}),
\end{align}
where $n$ is the total number of jumps, itself a random variable. A probability density of the trajectory compatible with the currents' statistics is given by
\begin{align}
p(\omega) = e^{- r_{x_{n}} \tau_{n}} \left( \prod_{i = 0}^{n-1} r_{x_{i+1}x_i} e^{- r_{x_i} \tau_{i}} \right) p_{t_0}(x_0)
\end{align}
where $r_x = \sum_{x'} r_{x'x}$ is the exit rate out of a state, and $p_{t_0}(x_0)$ is the distribution of the initial state. Currents are assumed to be anti-symmetric $\phi_a[\omega] = - \phi_a[\overline{\omega}]$ by time-reversal of the trajectory, defined as $\overline{\omega} := (x_{n},\tau_{n}) \to\ldots \to (x_1,\tau_1) \to (x_{0},\tau_{0})$. We focus on their mean and variance
\begin{align}
\label{eq:split}
\begin{split}
K^{(1)}_a(t_0,t) & := \langle \phi_a(t_0,t) \rangle \\
K^{(2)}_a(t_0,t) & := \langle \left( \phi_a(t_0,t) - \langle \,\phi_a(t_0,t) \rangle\right)^2 \rangle,
\end{split}
\end{align}
where $\langle\,\cdot\,\rangle$ is the expected value w.r.t. $p(\omega)$, and on their time-scaled versions $\kappa^{(i)}_a(t_0,t) =  K^{(i)}_a(t_0,t) / (t-t_0)$. The time scaling is introduced to account for the fact that all cumulants of the currents are time-extensive in the infinite-time limit, which in turn follows from the existence of a large deviation principle: while this is well-known for edge currents, for cycle currents this is established by Theorem 5 in \cite{jia16}. We are interested in particular in the time-averaged stationary mean and variance $\overline{\kappa}^{(k)}_a := \lim_{t \to \infty} \kappa^{(k)}_a(t_0,t)$, and the corresponding dispersion $\overline{\mathfrak{d}}_a := \overline{\kappa}^{(2)}/ \overline{\kappa}^{(1)}$, and in the istantaneous mean and variance $\kappa^{(k)}_a(t) := \lim_{dt \to 0} \kappa^{(k)}_a(t,t+dt)$, and the corresponding dispersion $\mathfrak{d}_a(t) := \kappa^{(2)}(t)/ \kappa^{(1)}(t)$.

The above edge TUR Eq.\,(\ref{eq:TUR}) is then established in terms of the transition forces $F_{x'x} := \log r_{x'x}/r_{xx'}$.


\paragraph{Cycle currents and involutions.} 

The first ingredient in our derivation is the decomposition of the trajectory $\omega$ as an ordered set of directed simple cycles $c \in \mathcal{C}$. For both cycle directions we introduce cycle fluxes $\psi_{\pm c}[\omega]$ and their antisymmetric part, the cycle currents $\phi_c[\omega] = \psi_{+c}[\omega] -  \psi_{-c}[\omega]$. One (of many) cycle decomposition of a trajectory follows the suggestion in Fig.\,\ref{fig:dedalus}. As the trajectory unfolds, we look at the first state that repeats itself, at transitions numbered $k$ and $k'$. Then the states $x_{k} \to x_{k+1} \to \ldots \to x_{k'}$ form a simple cycle $+c$:
\begin{equation}
 \ldots \to \stackrel{+c}{\overbrace{(x_k,\tau_k) \to (x_{k+1},\tau_{k+1}) \to \ldots \to (x_{k'} \equiv x_k ,\tau_{k'})}} \to \ldots 
\end{equation}
Every time one such cycle is identified we increase the corresponding cycle flux by one unit and then remove the corresponding transitions from the trajectory, yielding:
\begin{align}
 \ldots \to (x_{k'}, \tau_{k'}) \to \ldots
\end{align}
We proceed like this until we are left with a ``stump'', that is, a piece of trajectory from  $x_0$ to $x_{n}$ that contains no cycles. If the trajectory is closed, $x_0 = x_{n}$, then the stump consists of $(x_{n},\tau_{n})$ only.

We can now create a partial reversal of the trajectory by flipping the direction of cycle $\pm c$ into $\mp c$ whenever they occur, e.g.
\begin{equation}
\ldots \to \stackrel{-c}{\overbrace{(x_{k'},\tau_{k'}) \to (x_{k'-1},\tau_{k'-1}) \to \ldots \to (x_{k} ,\tau_{k})}} \to \ldots
\end{equation}
Proceeding in a similar manner for all cycles in a given family $c \in \mathcal{C}' \subseteq \mathcal{C}$ we obtain a new trajectory $\widetilde{\omega}$, that we call the partially reversed trajectory (see Fig.\ref{fig:dedalus}\,D). Now consider $p(\widetilde{\omega})$, where we sample the initial state with the same probability $p_{t_0}(x_0)$: in fact the initial state is the same for the forward and the partially reversed trajectory, as the ``stump'' is not affected by partial reversal. Also, the waiting-time distribution at states is exactly the same as in the forward trajectory. Finally, all transitions not belonging to the cycle will also be in the same direction. Therefore the following fluctuaton relation holds
\begin{align}
\frac{p(\omega)}{p(\widetilde{\omega})} = \exp \sum_{c \in \mathcal{C}'} F_c \, \phi_c[\omega], \label{eq:decon}
\end{align}
where we introduced the cycle affinity
\begin{align}
F_c  := \sum_{x'x \in c} F_{x'x} = \log \prod_{x'x \in c} \frac{r_{x'x}}{r_{xx'}}
\end{align}
and we used the obvious fact that all currents in the family are anti-symmetric by partial time-reversal. Importantly, the above fluctuation relation holds exactly at all times and does not require the long-time limit.

\paragraph{Exponential relation from Hilbert-space structure.}

The second crucial ingredient in our derivation is the Hilbert-space approach to uncertainties of Ref.\,\cite{fal20}. We consider the space $\mathcal{H}_{\mathcal{C}'}$ of square-integrable functions that are odd under partial time reversal $\omega  \to \widetilde{\omega}$, endowed with the scalar product $\langle f | g\rangle := \sum_{\omega } p(\omega ) f(\omega ) g(\omega )$. Defining $\widetilde{p}(\omega) := p(\widetilde{\omega})$, and using the antisymmetry, one finds that the observable $m = (p - \widetilde{p})/(p + \widetilde{p})$, living in the dual space $\mathcal{H}_{\mathcal{C}'}^\ast$, takes averages: $\langle f \rangle = \langle m | f \rangle$ for all $|f\rangle \in \mathcal{H}_{\mathcal{C}'}$. Then the variance of $f$ is $\langle f | f \rangle - \langle m|f\rangle^2$, and the Cauchy-Schwarz inequality $\langle m | f \rangle^2 \leq \langle m | m \rangle  \langle f | f \rangle$ yields
\begin{align}
\frac{\langle f | f \rangle - \langle f \rangle^2}{\langle f \rangle^2} \geq \frac{1}{\exp \langle s/2 \rangle -1},
\label{eq:exponentialbound}
\end{align}
where $s := \log p/\widetilde{p}$ and in the last inequality we used the (nontrivial) fact that $\langle \tanh s/2 \rangle \leq \tanh \langle s/2 \rangle$ \cite{fal20}. In view of the fluctuation relation Eq.\,(\ref{eq:decon}),  we find for an arbitrary linear combination $\phi_{\boldsymbol{a}}  = \sum_{c \in \mathcal{C}'} a_c \phi_c$ of observable cycle currents the exponential bound
{
\begin{align}
\frac{K^{(2)}_{\boldsymbol{a}}(t_0,t)}{K^{(1)}_{\boldsymbol{a}}(t_0,t)^2} \geq \frac{1}{\exp \tfrac{1}{2} \sum_{c \in \mathcal{C}'} F_c K^{(1)}_{c}(t_0,t) - 1}. \label{eq:fundamental}
\end{align}}

\paragraph{Istantaneous bound on cycle current.}

We are finally in the position to formulate our first main result. We consider short trajectories in the time interval $[t,t+dt)$. Because transition fluxes are linear combinations of cycle fluxes, and both are positive, and given that the former's average is of order $dt$, we know (as intuitive) that mean cycle currents are at most of order $dt$. Then we can linearize the exponential in Eq.\,(\ref{eq:exponentialbound}), and in the limit $dt \to 0$ we obtain
\begin{align}
\frac{\kappa^{(2)}_{\boldsymbol{a}}(t)}{\kappa^{(1)}_{\boldsymbol{a}}(t)^2} \geq \frac{2}{\sum_{c \in \mathcal{C}'} F_c \kappa^{(1)}_{c}(t)}.
\end{align}
In particular, selecting one particular cycle current $\boldsymbol{a} = c$, we arrive at the bound announced in Eq.\,(\ref{eq:cyclebound}). We investigate numerically the above inequality in the left-hand scatter plot of Fig.\,2 on one of the three simple cycles of the simple four-state model
 \begin{align}
\begin{array}{c}\xymatrix{1 \ar@{-}[r] & 2\ar@{-}[d] \\ 4 \ar@{-}[u]  \ar@{-}[ur] &  3 \ar@{-}[l]}  \end{array}; \qquad \mathcal{C} = \left\{
\Scale[.7]{\begin{array}{c}\xymatrix{ \ar@{-}[r] & \\  \ar@{-}[u]  \ar@{-}[ur] & }  \end{array}}, 
\Scale[.7]{\begin{array}{c}\xymatrix{  & \ar@{-}[d] \\ \ar@{-}[ur] &   \ar@{-}[l]}  \end{array}}, 
\Scale[.7]{\begin{array}{c}\xymatrix{ \ar@{-}[r] & \ar@{-}[d] \\  \ar@{-}[u] &   \ar@{-}[l]}  \end{array}}
 \right\}. \label{eq:model}
\end{align}
In the right-hand frame of Fig.\,2 we  further observe the validity of the long-time version of the bound.

\begin{figure}
\label{fig:random1}
\begin{center}
\includegraphics[width=0.51\columnwidth]{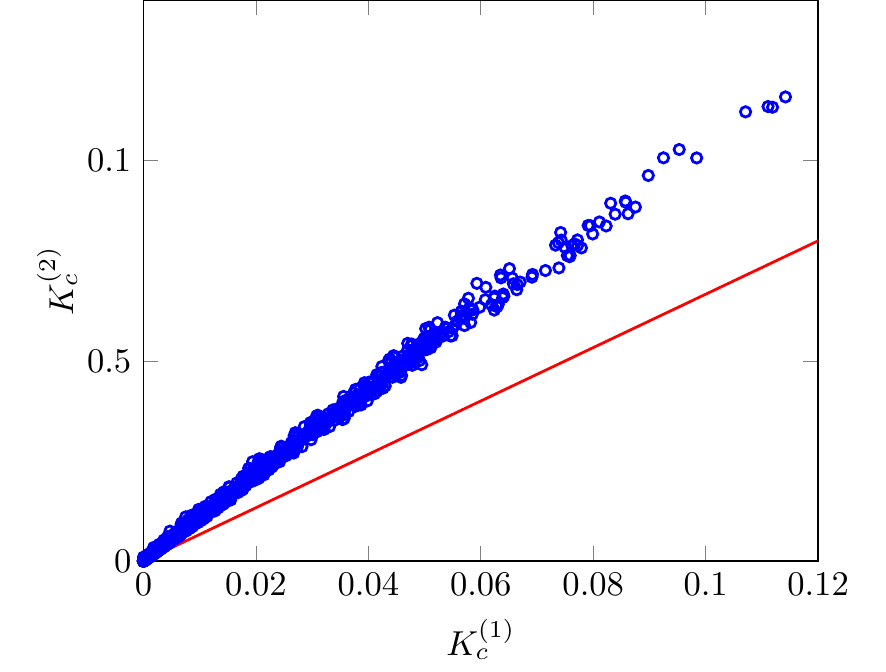}
\includegraphics[width=0.47\columnwidth]{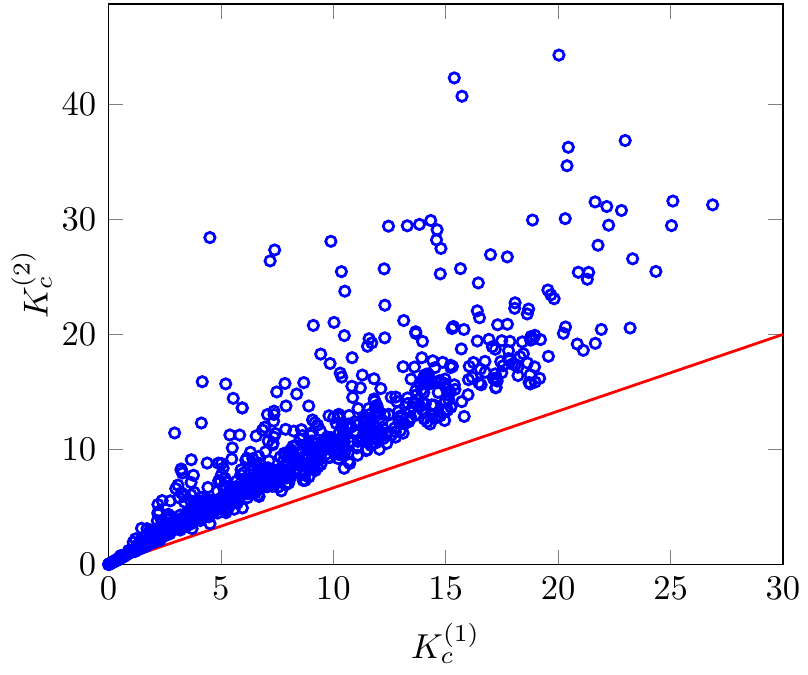}
\caption{Parametric plot of the mean and variance of the cycle current of cycle $c = 1 \to 2 \to 3 \to 1$ in the 4-state model depicted in Eq.\,(\ref{eq:model}). Data points are obtained via numerical simulation with the Gillespie algorithm. The current is obtained counting the net completion number of the cycle. Mean and variance are then calculated averaging over $10^4$ different realizations. The procedure is repeated for $10^3$ randomized systems with the transition rates $r_{x,x'}= s_{x,x'} e^{(u_x-u_{x'}+\frac 1 3 F_c)/2 } $, $s_{x,x'}=s_{x',x}$,  and $u_x$ uniformly distributed in $(0,1)$, corresponding to a cycle affinity $F_c=3$; the remaining transition rates are uniformly distributed in $(0,1)$. The trajectory duration is $|t-t_0|=1$ (left) and $|t-t_0|=10^2$ (right).} 
\end{center}
\end{figure}

\paragraph{Cycle bounds for transition currents.}

By construction, the number of times transition $x'x$ occurs equals the number of times some cycle through $x'x$ occurs. Therefore we have
\begin{align}
\phi_{x'x}(t_0,t) \approx \sum_{c \in \mathcal{C}_{x'x}}\phi_{c}(t_0,t) \label{eq:cyctran}
\end{align}
where $\mathcal{C}' = \mathcal{C}_{x'x}$ are all simple oriented cycles that contain transition $x'x$, and $\approx$ accounts for time-inextensive occurrences in the stump, which are of bounded variation. Plugging this latter equation into the entropy flow, and swapping the sum over transitions and that over cycles, we find as an important consistency check that cycle currents are thermodynamically consistent:
\begin{align}
\Sigma  \approx \sum_{x'< x} F_{x'x} \sum_{c \in   \mathcal{C}_{x'x}} \phi_c  = \sum_{c \in \mathcal{C}} F_{c} \phi_c. \label{eq:cycleEP}
\end{align}

Coming to our second main result, importantly transition currents are time-additive along trajectories, $\phi_{x'x}(t_0,t_2) = \phi_{x'x}(t_0,t_1) +  \phi_{x'x}(t_1,t_2)$ for $t_0 < t_1 < t_2$. 
This unlocks another argument in the derivation of Ref.\,\cite{fal20}, assuming that the system has already relaxed to a stationary state, $t_0 \to \infty$. Viewing this as a periodic state with period $\Delta t$, then the dispersion over an arbitrary number of periods $N =(t-t_0)/\Delta t $  is larger than the dispersion over a single period:
\begin{align}
\frac{K^{(2)}_{x'x}(t_0,t_0+N \Delta t)}{K^{(1)}_{x'x}(t_0,t_0+ N \Delta t)} \geq  \frac{K^{(2)}_{x'x}(t_0,t_0+ \Delta t)}{K^{(1)}_{x'x}(t_0, t_0+\Delta t)}.
\label{eq:periodic}
\end{align}
We now let $\Delta t \to 0$. Defining $\sigma_{x'x} := \sum_{c \in \mathcal{C}_{x'x} } F_c \overline{\kappa}^{(1)}_c$, and given that $\kappa^{(1)}_{\boldsymbol{a}}(t) = \overline{\kappa}^{(1)}_{\boldsymbol{a}} $ thanks to stationarity, we can use Eq.\,\eqref{eq:fundamental} with $\boldsymbol{a}=x'x$ to bound the right-hand side of \eqref{eq:periodic}, leading to
\begin{align}
\overline{\mathfrak{d}}_{x'x}  \geq   \frac{2 }{ \sigma_{x'x} / |\overline{\kappa}^{(1)}_{x'x} | }.
\label{eq:TURtran}
\end{align}

\begin{figure}[t!]
  \centering
  \includegraphics[width=.9\columnwidth]{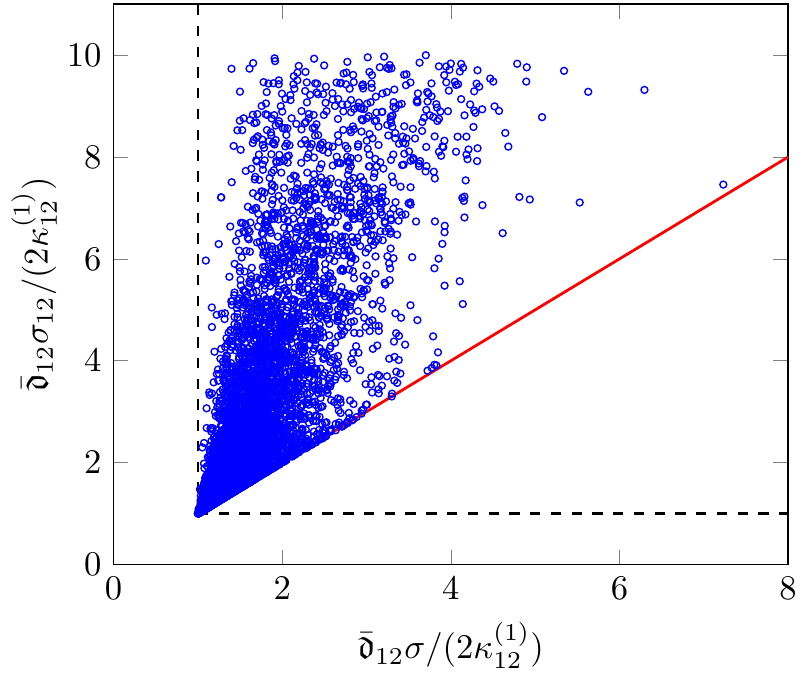}
\caption{Scatter plot of $\overline{\mathfrak{d}}_{12}\sigma_{12}/(2\kappa^{(1)}_{12})$ in terms of $\overline{\mathfrak{d}}_{12}\sigma/(2\kappa^{(1)}_{12})$ for systems with randomized rates in the unit interval, showing that both bounds are satisfied (all points are above the $x=1$, $y=1$ axes), and that the local bound performs better than the global one (all points are above the $x=y$ line).}
\label{fig:rate}
\end{figure}

Let us now prove that this bound improves on the global one. To compute $\sigma_{x'x}$, we use a known  \cite{jia04, zia07} analytical expression for the mean stationary cycle currents as $\overline{\kappa}^{(1)}_c = S_c \left(P^+_c - P^-_c \right)$. Here, $P^\pm_c$ are respectively the products of rates in clockwise/counterclockwise directions along the cycle, while $S_c$ is a positive factor, symmetric by reversal of the cycle \cite{zia07}. Because $F_c = \log P^+_c/P^-_c$ and $(x-y) \log x/y \geq 0$, we find that each term in $\sigma_{x'x}$ is non-negative. Furthermore, given Eq.\,(\ref{eq:cycleEP}), because we are summing over a subset of all simple cycles, we have that $\sigma_{x'x} \leq \sigma$. We illustrate this result in Fig.\,\ref{fig:rate}.


\paragraph{Discussion.}

While extensive in time, cycle currents are not additive. This is already evident from our illustration in Fig.\,1, where one of the later cycles recorded by Theseus actually initiated earlier than another. For this reason, several results known for edge currents do not immediately apply to cycle currents. In particular we were not able to prove the long-time averaged uncertainty relation. Regarding the simulations sustaining it (see right panel of Fig.\,2), while we were cautious about self-correlation and relaxation errors already present in MCMC algorithms \cite{sok97}, due to their nonlocal correlations cycle currents may pose specific systematic errors that need to be investigated further.


As regards the improved bound on transition currents, as the system size grows, the number of cycles containing one particular transition grows much slower than the total number of cycles. For example, in a complete graph with $V$ vertices there are $\sum_{k = 3}^V V!/(V-k)! 2 k$ simple cycles, while the number of cycles through a particular edge (not counting the trivial cycle) is $\floor*{(V-2)! e} - 1$: for the first few values of $V \geq 3$ the ratio of local-to-global cycles is $1$, $4/7$, $15/37$, $64/197$, $325/1172$, $978/4009$. In more general cases the number $n(C)$ of simple cycles for a graph with cyclomatic number $C = E - V + 1$ (edges minus vertices plus one) is $2^{C}-1 \geq n(C) \geq 2^{C-1} +C^2 - 3C + 3$, and usually the lower bound is a good approximation \cite{ent81}. To the best of our knowledge, estimates on the number of cycles sharing a given edge are not known, but since a cyclomatic number of simple basis cycles is sufficient to compose any simple cycle, and since the basis cycles that compose a given simple cycle must be adjacent one to another, then simple cycles could be viewed as walks in the dual graph/matroid, and such estimates may be mapped into known walk-enumeration problems. Finally, when considering not just the bare cycle number, but the dissipation each cycle provides, assuming the rates to be homogeneously distributed over the graph, factor $S_c$ has a tendency to become smaller the larger cycle $c$ is, that is, the further away it goes from the rooting vertex, due to the fact that this factor measures the contraction of the number of spanning trees upon identification of the cycle with a unique vertex \cite{zia07}.

\paragraph{Conclusions.}

All of this indicates that in larger systems local cycle bounds on edge currents may perform enormously better than global ones. Notice the give-and-get: in order to go local in results, we have to consider an intermediate, less local representation of the observable.

The question left open is then about the physical relevance of cycle currents. In this respect, a conceptual shift may be needed about how we conceive of resources. Single transitions are associated to transfer of given amounts of matter, energy etc. (i.e. an entropy change in the reservoirs that affect that transition). When moving to cycle currents, the way in which transitions follow in time matters. Such approaches may be relevant when considering the thermodynamics of processes which need to go through an ordered sequence of events to be completed, in analogy to a product that has to go through different stages of production along a factory line \cite{pan17}. Such situations seem to arise at the cellular level when a cell needs to undergo a well defined sequence of transformations before dividing \cite{ahm20}. However, with the exception of Ref.\,\cite{jia16} and previous work by the same Authors, little systematic effort has been made to actually establish cycles as the grounding point of more advanced thermodynamic analysis, e.g. by developing {\it ad hoc} perturbative methods or algorithms.

\paragraph*{Acknowledgments.} 

The research was supported by the National Research Fund Luxembourg (project CORE ThermoComp R-AGR-3425-10) and by the European Research Council, project NanoThermo (ERC-2015-CoG Agreement No. 681456).

\bibliography{biblio}

\end{document}